\documentclass{sig-alternate}

\usepackage{algorithm,color}
\usepackage[noend]{algpseudocode}
\usepackage{multirow}
\usepackage{rotating}
\usepackage{array}
\usepackage[protrusion=true,final]{microtype}
\newtheorem{theorem}{Theorem}[section]

\newcommand{\lt}{\left}
\newcommand{\rt}{\right}
\newcommand\T{\rule{0pt}{2.6ex}}
\newcommand\B{\rule[-1.6ex]{0pt}{0pt}}

\graphicspath{{fig/}}

\begin{document}\sloppy

\clubpenalty=10000
\widowpenalty = 10000

\title{Communication-Optimal Parallel Algorithm for \\ Strassen's Matrix Multiplication\\
}

\numberofauthors{5}
\author{
\alignauthor
Grey Ballard\titlenote{Research supported by Microsoft (Award $\#$024263) and Intel (Award $\#$024894) funding and by matching funding by U.C. Discovery (Award $\#$DIG07-10227). Additional support comes from Par Lab affiliates National Instruments, Nokia, NVIDIA, Oracle, and Samsung.}\\
\affaddr{EECS Department}\\
\affaddr{UC Berkeley}\\
\affaddr{Berkeley, CA 94720}\\
\email{ballard@eecs.berkeley.edu}
\alignauthor
James Demmel$^{^{^{\textrm{\normalsize $*$}}}}$\titlenote{Research is also supported by DOE grants DE-{}SC0003959, DE-{}SC0004938, and DE-AC02-05CH11231.}\\
\affaddr{Mathematics Department}\\
\affaddr{and CS~Division}\\
\affaddr{UC Berkeley}\\
\affaddr{Berkeley, CA 94720}\\
\email{demmel@cs.berkeley.edu}
\alignauthor
Olga Holtz\titlenote{Research supported by the Sofja Kovalevskaja
programme of Alexander von Humboldt Foundation
and by the National Science Foundation under agreement DMS-0635607,
while visiting the Institute for Advanced Study.}\\
\affaddr{Mathematics Departments}\\
\affaddr{UC Berkeley}\\
\affaddr{and TU Berlin}\\
\affaddr{Berkeley, CA 94720}\\
\email{holtz@math.berkeley.edu}
  \and
Benjamin Lipshitz$^{^{^{\textrm{\normalsize $*$}}}}$\\
\affaddr{EECS Department}\\
\affaddr{UC Berkeley}\\
\affaddr{Berkeley, CA 94720}\\
\email{ lipshitz@berkeley.edu}
\alignauthor
Oded Schwartz\titlenote{
Research supported by U.S. Department of Energy grants under
Grant Numbers DE-{}SC0003959.
}\\
\affaddr{EECS Department}\\
\affaddr{UC Berkeley}\\
\affaddr{Berkeley, CA 94720}\\
\email{odedsc@eecs.berkeley.edu}
} 

\maketitle

\begin{abstract}
Parallel matrix multiplication is one of the most studied fundamental problems in distributed and high performance computing. We obtain a new parallel algorithm that is based on Strassen's fast matrix multiplication and minimizes communication. The algorithm outperforms all known parallel matrix multiplication algorithms, classical and Strassen-based, both asymptotically and in practice.

A critical bottleneck in parallelizing Strassen's algorithm is the communication between the processors. Ballard, Demmel, Holtz, and Schwartz (SPAA'11) prove lower bounds on these communication costs, using expansion properties of the underlying computation graph. Our algorithm matches these lower bounds, and so is communication-optimal. It exhibits perfect strong scaling within the maximum possible range.  

Benchmarking our implementation on a Cray XT4, we obtain speedups over classical and Strassen-based algorithms ranging from $24\%$ to $184\%$ for a fixed matrix dimension $n=94080$, where the number of nodes ranges from $49$ to $7203$.

Our parallelization approach generalizes to other fast matrix multiplication algorithms.
\end{abstract}

\smallskip \noindent
{\bf Categories and Subject Descriptors}: F.2.1 [Analysis of Algorithms and Problem Complexity]: Numerical Algorithms and Problems: Computations on matrices \\
{\bf ACM General Terms}: algorithms 

\twocolumn
\section{Introduction}

Matrix multiplication is one of the most fundamental algorithmic problems in numerical linear algebra, distributed computing,
scientific computing, and high-performance computing. Parallelization of matrix multiplication has been extensively studied ({\em e.g.}, \cite{Cannon69,Berntsen89,ChoiDongarraWalker93,AgarwalBalleGustavsonJoshiPalkar95,LD95,GSvdG95,SUMMA,Choi98,IronyToledoTiskin04,SolomonikDemmel11,BallardDemmelHoltzSchwartz11b,BallardDemmelHoltzLipshitzSchwartz12a}). It has been addressed using many theoretical approaches, algorithmic tools, and software engineering methods in order to optimize performance and obtain faster and more efficient parallel algorithms and implementations.

We obtain a new parallel algorithm based on Strassen's fast matrix multiplication.\footnote{Our actual implementation uses the Winograd variant \cite{Winograd71}; see Appendix~\ref{app:str-win} for details.} It is more efficient than any other parallel matrix multiplication algorithm of which we are aware, including those that are based on classical (\(\Theta(n^3)\)) multiplication, and those that are based on Strassen's and other Strassen-like matrix multiplications. We compare the efficiency of the new algorithm with previous algorithms, and provide both asymptotic analysis (Sections~\ref{sec:algorithm} and~\ref{sec:parallelalgforStrassen}) and benchmarking data (Section \ref{sec:performance}).

\subsection{The communication bottleneck} 
To design efficient parallel algorithms, it is necessary not only to load balance the computation, but also to minimize the time spent communicating between processors. The inter-processor communication costs are in many cases significantly higher than the computational costs. Moreover, hardware trends predict that more problems will become communication-bound in the future \cite{GrahamSnirPatterson04,FullerMillett11}.  Even matrix multiplication becomes communication-bound when run on sufficiently many processors.  Given the importance of communication costs, it is preferable to match the performance of an algorithm to a communication lower bound, obtaining a communication-optimal algorithm.

\subsection{Communication costs of matrix multiplication}
We consider a distributed-memory parallel machine model as described in Section~\ref{sec:model}.  The communication costs are measured as a function of the number of processors $P$, the local memory size $M$ in words, and the matrix dimension \(n\). Irony, Toledo, and Tiskin \cite{IronyToledoTiskin04} proved that in the distributed memory parallel model, the bandwidth cost of classical $n$-by-$n$ matrix multiplication is bounded by $\Omega \lt(\frac{n^3}{PM^{1/2}}\rt)$ words. Using their technique one can also deduce a memory-independent bandwidth cost bound of $\Omega \lt(\frac{n^2}{P^{2/3}}\rt)$\cite{BallardDemmelHoltzLipshitzSchwartz12a} and generalize it to other classes of algorithms \cite{BallardDemmelHoltzSchwartz11a}. For a shared-memory model similar bounds were shown in \cite{AggarwalChandraSnir90}. Until recently, parallel classical matrix multiplication algorithms (e.g., ``2D'' \cite{Cannon69,SUMMA}, and ``3D'' \cite{Berntsen89,AgarwalBalleGustavsonJoshiPalkar95}) have minimized communication only for specific $M$ values.  The first algorithm that minimizes the communication costs for the entire range of $M$ has recently been obtained by Solomonik and Demmel \cite{SolomonikDemmel11}.  See Section~\ref{sec:classical_algs} for more details.

None of these lower bounding techniques and parallelization approaches generalize to fast matrix multiplication, such as
\cite{Strassen69,Pan80,Bini80,Schonhage81,Romani82,CoppersmithWinograd82,Strassen87,CoppersmithWinograd87,CohnKleinbergSzegedyUmans05,VassilevskaWilliams12}. A communication cost lower bound for fast matrix multiplication algorithms has only recently been obtained \cite{BallardDemmelHoltzSchwartz11b}: Strassen's algorithm run on a distributed-memory parallel machine has bandwidth cost $\Omega \lt(\lt(\frac{n}{M^{1/2}}\rt)^{\omega_0}\cdot \frac MP \rt)$  and latency cost $\Omega \lt(\lt(\frac{n}{M^{1/2}}\rt)^{\omega_0}\cdot \frac 1P \rt)$, where $\omega_0 = \log_2 7$ (see Section~\ref{sec:LB}). These bounds generalize to other, but not all, fast matrix multiplication algorithms, with $\omega_0$ being the exponent of the computational complexity.  

In the sequential case,\footnote{See \cite{BallardDemmelHoltzSchwartz11b} for a discussion of the sequential memory model.} the lower bounds are attained by the natural recursive implementation \cite{Strassen69} which is thus optimal. However, a parallel communication-optimal Strassen-based algorithm was not previously known. Previous parallel algorithms that use Strassen ({\em e.g.}, \cite{GSvdG95,LD95,DesprezSuter04}), decrease the computational costs at the expense of higher communication costs.  The factors by which these algorithms exceed the lower bounds are typically small powers of \(P\) and \(M\), as discussed in Section~\ref{sec:otheralg}.  However both \(P\) and \(M\) can be large ({\em e.g.} on a modern supercomputer, one may have \(P\sim 10^5\) and \(M\sim 10^9\)).

\subsection{Parallelizing Strassen's matrix multiplication in a communication efficient way}

The main impetus for this work was the observation of the asymptotic gap between the communication costs of existing parallel Strassen-based algorithms and the communication lower bounds.  Because of the attainability of the lower bounds in the sequential case, we hypothesized that the gap could be closed by finding a new algorithm rather than by tightening the lower bounds.  

We made three observations from the lower bound results of \cite{BallardDemmelHoltzSchwartz11b} that lead to the new algorithm.  First, the lower bounds for Strassen are lower than those for classical matrix multiplication.  This implies that in order to obtain an optimal Strassen-based algorithm, the communication pattern for an optimal algorithm cannot be that of a classical algorithm but must reflect the properties of Strassen's algorithm.  Second, the factor $M^{\omega_0/2-1}$ that appears in the denominator of the communication cost lower bound implies that an optimal algorithm must use as much local memory as possible.  That is, there is a tradeoff between memory usage and communication (the same is true in the classical case). Third, the proof of the lower bounds shows that in order to minimize communication costs relative to computation, it is necessary to perform each sub-matrix multiplication of size  $\Theta(\sqrt M) \times \Theta(\sqrt M)$ on a single processor.

With these observations and assisted by techniques from previous approaches to parallelizing Strassen, we developed a new parallel algorithm which achieves perfect load balance, minimizes communication costs, and in particular performs asymptotically less computation and communication than is possible using classical matrix multiplication.

\subsection{Our contributions and paper organization}
Our main contribution is a new algorithm we call Communication-Avoiding Parallel Strassen, or {\em CAPS}.
\begin{theorem}
\label{thm:main1}
CAPS asymptotically minimizes computational and bandwidth costs over all parallel Strassen-based algorithms.  It also minimizes latency cost up to a logarithmic factor in the number of processors.
\end{theorem}
CAPS performs asymptotically better than any previous previous classical or Strassen-based parallel algorithm. It also runs faster in practice. The algorithm and its computational and communication cost analyses are presented in Section~\ref{sec:algorithm}. There we show it matches the communication lower bounds. 

We provide a review and analysis of previous algorithms in Section~\ref{sec:otheralg}. We also consider two natural combinations of previously known algorithms (Sections~\ref{sec:25d-strassen} and~\ref{sec:strassen-25d}). One of these new algorithms that we call ``2.5D-Strassen'' performs better than all previous algorithms, but is still not optimal, and performs worse than CAPS.

We discuss our implementations of the new algorithms and compare their performance with previous ones in Section~\ref{sec:performance} to show that our new CAPS algorithm outperforms previous algorithms not just asymptotically, but also in practice.  Benchmarking our implementation on a Cray XT4, we obtain speedups over classical and Strassen-based algorithms ranging from $24\%$ to $184\%$ for a fixed matrix dimension $n=94080$, where the number of nodes ranges from $49$ to $7203$.

In Section~\ref{sec:conclusion} we show that our parallelization method applies to other fast matrix multiplication algorithms. It also applies to classical recursive matrix multiplication, thus obtaining a new optimal classical algorithm that matches the 2.5D algorithm of Solomonik and Demmel \cite{SolomonikDemmel11}. In Section~\ref{sec:conclusion}, we also discuss numerical stability, hardware scaling, and future work.

\section{Preliminaries}
\label{sec:prelim}
\subsection{Communication model}
\label{sec:model}

We model communication of distributed-memory parallel architectures as follows.  We assume the machine has \(P\) processors, each with local memory of size $M$ words, which are connected via a network.  Processors communicate via messages, and we assume that a message of $w$ words can be communicated in time $\alpha+\beta w$.  The bandwidth cost of the algorithm is given by the word count and denoted by \(BW(\cdot)\), and the latency cost is given by the message count and denoted by \(L(\cdot)\).  Similarly the computational cost is given by the number of floating point operations and denoted by \(F(\cdot)\). We call the time per floating point operation \(\gamma\).

We count the number of words, messages and floating point operations along the \emph{critical path} as defined in \cite{YangMiller88}.  That is, two messages that are communicated between separate pairs of processors simultaneously are counted only once, as are two floating point operations performed in parallel on different processors.  This metric is closely related to the total running time of the algorithm, which we model as
\[\alpha L(\cdot)+\beta BW(\cdot)+\gamma F(\cdot).\]

We assume that (1) the architecture is homogeneous (that is, \(\gamma\) is the same on all processors and \(\alpha\) and \(\beta\) are the same between each pair of processors), (2) processors can send/receive only one message to/from one processor at a time and they cannot overlap computation with communication (this latter assumption can be dropped, affecting the running time by a factor of at most two), and (3) there is no communication resource contention among processors. That is, we assume that there is a link in the network between each pair of processors.  Thus lower bounds derived in this model are valid for any network, but attainability of the lower bounds depends on the details of the network.
\subsection{Strassen's algorithm}

Strassen showed that \(2\times 2\) matrix multiplication can be performed using \(7\) multiplications and \(18\) additions, instead of the classical algorithm that does \(8\) multiplications and \(4\) additions \cite{Strassen69}.  By recursive application this yields an algorithm with multiplies two \(n\times n\) matrices  \(O(n^{\omega_0})\) flops, where \(\omega_0=\log_2 7\approx 2.81\).  Winograd improved the algorithm to use \(7\) multiplications and \(15\) additions in the base case, thus decreasing the hidden constant in the \(O\) notation \cite{Winograd71}.  We review the Strassen-Winograd algorithm in Appendix~\ref{app:str-win}.

\subsection{Previous work on parallel Strassen}

In this section we breifly describe previous efforts to parallelize Strassen.  More details, including communication analyses, are in Section~\ref{sec:parallelalgforStrassen}.  A summary appears in Table~\ref{tbl:summary}.

Luo and Drake \cite{LD95} explored Strassen-based parallel algorithms that use the communication patterns known for classical matrix multiplication.  They considered using a classical 2D parallel algorithm and using Strassen locally, which corresponds to what we call the ``2D-Strassen'' approach (see Section~\ref{sec:alg:can-str}).  They also consider using Strassen at the highest level and performing a classical parallel algorithm for each subproblem generated, which corresponds to what we call the ``Strassen-2D'' approach.  The size of the subproblems depends on the number of Strassen steps taken (see Section~\ref{sec:alg:str-can}).  Luo and Drake also analyzed the communication costs for these two approaches.

Soon after, Grayson, Shah, and van de Geijn \cite{GSvdG95} improved on the Strassen-2D approach of \cite{LD95} by using a better classical parallel matrix multiplication algorithm and running on a more communication-efficient machine.  They obtained better performance results compared to a purely classical algorithm for up to three levels of Strassen's recursion.

Kumar, Huang, Johnson, and Sadayappan \cite{KHJS93} implemented Strassen's algorithm on a shared-memory machine.  They identified the tradeoff between available parallelism and total memory footprint by differentiating between ``partial'' and ``complete'' evaluation of the algorithm, which corresponds to what we call depth-first and breadth-first traversal of the recursion tree (see Section~\ref{sec:CAPS-overview}).  They show that by using $\ell$ DFS steps before using BFS steps, the memory footprint is reduced by a factor of $(7/4)^\ell$ compared to using all BFS steps.  They did not consider communication costs in their work.

Other parallel approaches \cite{DesprezSuter04,HunoldRauberRunger08,SDM06} have used more complex parallel schemes and communication patterns.  However, they restrict attention to only one or two steps of Strassen and obtain modest performance improvements over classical algorithms.

\subsection{Strassen lower bounds}
\label{sec:LB}

For Strassen-based algorithms, the bandwidth cost lower bound has been proved using expansion arguments on the computation graph, and the latency cost lower bound is an immediate corollary.

\begin{theorem} (Memory-dependent lower bound) \cite{BallardDemmelHoltzSchwartz11b}
\label{thm:mem-dep-lb} 
Consider a Strassen-based algorithm running on \(P\) processors each with local memory size \(M\).  Let \(BW(n,P,M)\) be the bandwith cost and \(L(n,P,M)\) be the latency cost of the algorithm.  Assume that no intermediate values are computed twice.  Then
\[BW(n,P,M) = \Omega\lt( \lt(\frac{n}{\sqrt M } \rt)^{\omega_0}\cdot \frac{M}{P} \rt),\]
\[L(n,P,M) = \Omega\lt( \lt(\frac{n}{\sqrt M } \rt)^{\omega_0}\cdot \frac{1}{P} \rt).\]
\end{theorem}

A memory-independent lower bound has recently been proved using the same expansion approach:

\begin{theorem} (Memory-independent lower bound) \cite{BallardDemmelHoltzLipshitzSchwartz12a}
\label{thm:mem-indep-lb}
Consider a Strassen-based algorithm running on \(P\) processors.  Let \(BW(n,P)\) be the bandwith cost and \(L(n,P)\) be the latency cost of the algorithm.  Assume that no intermediate values are computed twice.  Assume only one copy of the input data is stored at the start of the algorithm and the computation is load-balanced in an asymptotic sense.  Then
\[BW(n,P) = \Omega\lt( \frac{n^2}{P^{2/\omega_0}} \rt),\]
and the latency cost is \(L(n,P)=\Omega(1)\).
\end{theorem}

Note that when \(M=O(n^2/P^{2/\omega_0})\), the memory-dependent lower bound is dominant, and when \(M=\Omega(n^2/P^{2/\omega_0})\), the memory-independent lower bound is dominant.

\section{Communication-Avoiding\\Parallel Strassen}

In this section we present the CAPS algorithm, and prove it is communication-optimal.  See Algorithm~\ref{alg:brief} for a concise presentation and Algorithm~\ref{alg:detail} for a more detailed description.
\begin{theorem}
\label{thm:main2}
CAPS has computational cost \(\Theta\lt(\frac{n^{\omega_0}}{P}\rt)\), bandwidth cost \(\Theta\lt ( \max \lt\{
\frac{n^{\omega_0}}{PM^{\omega_0/2-1}},\frac{n^2}{P^{2/\omega_0}}\rt\}\rt)\), and latency cost
\(\Theta\lt(\max\lt\{\frac{n^{\omega_0}}{PM^{\omega_0/2}}\log
P,\log P\rt\}\rt)\).
\end{theorem}
By Theorems~\ref{thm:mem-dep-lb} and~\ref{thm:mem-indep-lb}, we see that CAPS has optimal computational and bandwidth costs, and that its latency cost is at most \(\log P\) away from optimal.  Thus Theorem~\ref{thm:main1} follows.  We prove Theorem~\ref{thm:main2} in Section~\ref{sec:algorithm:optimal}.

\label{sec:algorithm}

\subsection{Overview of CAPS}
\label{sec:CAPS-overview}

\begin{figure}[t!]
  \centering
  \includegraphics[width=3.3in]{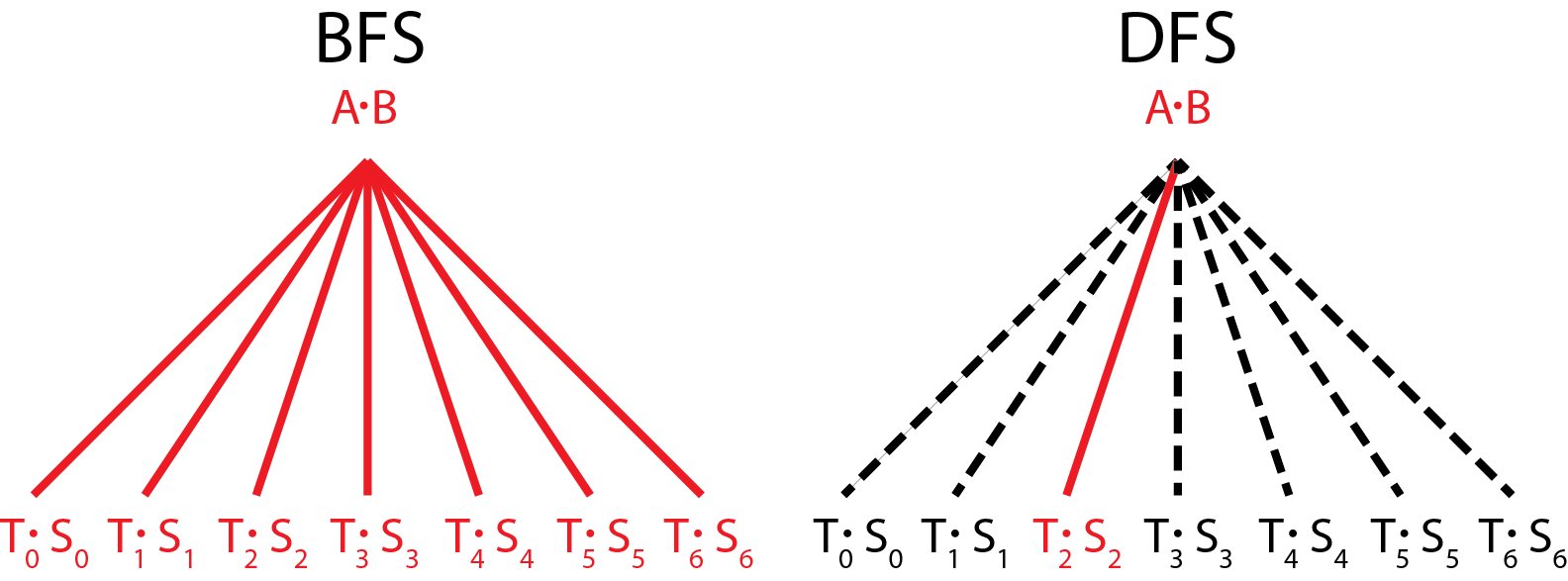}
  \caption{Representation of BFS and DFS steps.  In a BFS step, all seven subproblems are computed at once, each on 1/7 of the processors.  In a DFS step, the seven subproblems are computed in sequence, each using all the processors.  The notation follows that of Appendix~\ref{app:str-win}.}
  \label{fig:bfsdfs}
\end{figure}

Consider the recursion tree of Strassen's sequential algorithm.  CAPS traverses it in parallel as follows.  At each level of the tree, the algorithm proceeds in one of two ways.  A ``breadth-first-step'' {\em (BFS)} divides the 7 subproblems among the processors, so that $\frac17$ of the processors work on each subproblem independently and in parallel.  A ``depth-first-step'' {\em (DFS)} uses all the processors on each subproblem, solving each one in sequence.  See Figure~\ref{fig:bfsdfs}.

In short, a BFS step requires more memory but reduces communication costs while a DFS step requires little extra memory but is less communication-efficient.  In order to minimize communication costs, the algorithm must choose an ordering of BFS and DFS steps that uses as much memory as possible.

Let \(k=\log_7 P\) and \(s\geq k\) be the number of distributed Strassen steps the algorithm will take.  In this section, we assume that \(n\) is a multiple of \(2^s7^{\lceil k/2\rceil}\).  If \(k\) is even, the restriction simplifies to \(n\) being a multiple of \(2^s\sqrt P\). Since \(P\) is a power of 7, it is sometimes convenient to think of the processors as numbered in base 7. CAPS performs \(s\) steps of Strassen's algorithm and finishes the calculation with local matrix multiplication.  The algorithm can easily be generalized to other values of \(n\) by padding or dynamic peeling.

We consider two simple schemes of traversing the recursion tree with BFS and DFS steps.  The first scheme, which we call the Unlimited Memory {\em (UM)} scheme, is to take $k$ BFS steps in a row.  This approach is possible only if there is sufficient available memory.  The second scheme, which we call the Limited Memory {\em (LM)} scheme is to take $\ell$ DFS steps in a row followed by $k$ BFS steps in a row, where $\ell$ is minimized subject to the memory constraints.

It is possible to use a more complicated scheme that interleave BFS and DFS steps to reduce communication.  We show that the LM scheme is optimal up to a constant factor, and hence no more than a constant factor improvement can be attained from interleaving.

\begin{algorithm}
\caption{CAPS, in brief. For more details, see Algorithm~\ref{alg:detail}.}
\label{alg:brief}
\begin{algorithmic}[1]
\Require $A$, $B$, $n$, where $A$ and $B$ are \(n\times n\) matrices
\Statex\hskip.6cm $P =$ number of processors
\Ensure $C=A\cdot B$
\Procedure{C = CAPS}{$A$, $B$, $n$, $P$}
\If{enough memory}\Comment{Do a BFS step}
  \State locally compute the $S_i$'s and $T_i's$ from $A$ and $B$ 
  \While{$i = 1\ldots 7$}
    \State redistribute $S_i$ and $T_i$
    \State $Q_i=$ CAPS($S_i$, $T_i$, $n/2$, $P/7$)
    \State redistribute $Q_i$
  \EndWhile
  \State locally compute $C$ from all the $Q_i$'s 
\Else
  \Comment{Do a DFS step}
  \For{$i = 1\ldots 7$}
    \State locally compute $S_i$ and $T_i$ from A and B 
    \State $Q_i=$ CAPS($S_i$, $T_i$, $n/2$, $P$)
    \State locally compute contribution of $Q_i$ to $C$ 
  \EndFor
\EndIf
\EndProcedure
\Statex\Comment{The dependence of the \(S_i\)'s on \(A\), the \(T_i\)'s on \(B\) and \(C\) on the \(Q_i\)'s follows the Strassen or Strassen-Winograd algorithm.  See Appendix~\ref{app:str-win}.}
\end{algorithmic}
\end{algorithm}

\subsection{Data layout}
\label{sec:datalayout}
\begin{figure}[h!]
  \centering
  \includegraphics[width=3.3in]{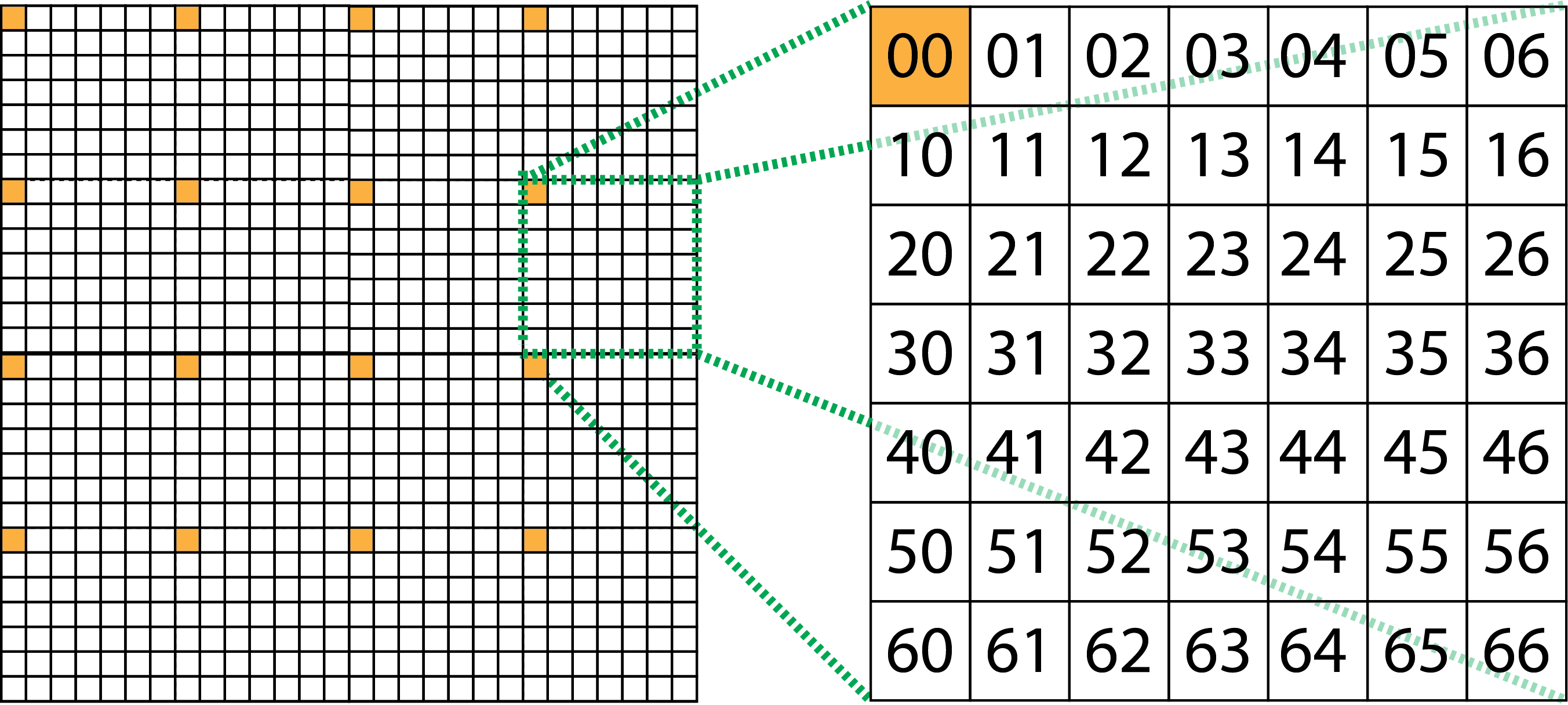}
  \caption{An example matrix layout for CAPS.  Each of the 16 submatrices as shown on the left has exactly the same layout.  The colored blocks are the ones owned by processor 00. On the right is a zoomed-in view of one submatrix, showing which processor, numbered base 7, owns each block.  This is block-cyclic layout with some blocksize \(b\), and matches our layout requirements with parameters \((n=4\cdot49\cdot b,P=49,s=2)\).}
  \label{fig:layout}
\end{figure}

We require that the data layout of the matrices satisfies the following two properties:
\begin{enumerate}
\item At each of the \(s\) Strassen recursion steps, the data layouts of the four sub-matrices of each of \(A\), \(B\), and \(C\) must match so that the weighted additions of these sub-matrices can be performed locally.  This technique follows \cite{LD95} and allows communication-free DFS steps.
\item Each of these submatrices must be equally distributed among the \(P\) processors for load balancing.
\end{enumerate}
There are many data layouts that satisfy these properties, perhaps the simplest being block-cyclic layout with a processor grid of size \(7^{\lfloor k/2 \rfloor} \times 7^{\lceil k/2 \rceil}\) and block size \(\frac{n}{2^s7^{\lfloor k/2 \rfloor}}\times\frac{n}{2^s7^{\lceil k/2 \rceil}}\). (When \(k=\log_7 P\) is even these expressions simplify to a processor grid of size \(\sqrt{P}\times\sqrt{P}\) and block size \(\frac{n}{2^s\sqrt{P}}\).)  See Figure~\ref{fig:layout}.

Any layout that we use is specified by three parameters, \((n,P,s)\), and intermediate stages of the computation use the same layout with smaller values of the parameters.  A BFS step reduces a multiplication problem with layout parameters \((n,P,s)\) to seven subproblems with layout parameters \((n/2,P/7,s-1)\).  A DFS step reduces a multiplication problem with layout parameters \((n,P,s)\) to seven subproblems with layout parameters \((n/2,P,s-1)\).

Note that if the input data is initially load-balanced but distributed using a different layout, we can rearrange it to the above layout using a total of \(O\lt(\frac{n^2}{P}\rt)\) words and \(O(n^2)\) messages.  This has no asymptotic effect on the bandwidth cost but significantly increases the latency cost in the worst case.

\begin{algorithm}[ht!]
\caption{CAPS, in detail}
\label{alg:detail}
\begin{algorithmic}[1]
\Require $A$, $B$, are \(n\times n\) matrices
\Statex\hskip.6cm \(P=\) number of processors
\Statex\hskip.6cm rank $=$ processor number base-7 as an array
\Statex\hskip.6cm $M=$ local memory size
\Ensure $C=A\cdot B$
\Procedure{C = CAPS}{$A$, $B$, $P$, rank, $M$}
\State $\ell = \left\lceil\log_2 \frac{4n}{P^{1/{\omega_0}}M^{1/2}}\right\rceil$ 
\Statex \Comment{\(\ell\) is number of DFS steps to fit in memory}
\State $k=\log_7 P$
\State {\bf call} DFS($A$, $B$, $C$, $k$, $\ell$, rank)
\EndProcedure
\end{algorithmic}
\line(1,0){240}
\begin{algorithmic}[1]
\Procedure{DFS}{$A$, $B$, $C$, $k$, $\ell$, rank}
\Statex\Comment{Do $C=A\cdot B$ by \(\ell\) DFS, then \(k\) BFS steps}
  \If{$\ell \leq 0$}
    {\bf call} BFS( $A$, $B$, $C$, $k$, rank);
     {\bf return}
  \EndIf
  \For{$i = 1\ldots 7$}
    \State locally compute $S_i$ and $T_i$ from $A$ and $B$ \Statex\Comment{following Strassen's algorithm}
    \State {\bf call} DFS( $S_i$, $T_i$, $Q_i$, $k$, $\ell-1$, rank )
    \State locally compute contribution of $Q_i$ to $C$ \Statex\Comment{following Strassen's algorithm}
  \EndFor
\EndProcedure
\end{algorithmic}
\line(1,0){240}
\begin{algorithmic}[1]
\Procedure{BFS}{$A$, $B$, $C$, $k$, rank}
\Statex\Comment{Do $C=A\cdot B$ by k BFS steps, then local Strassen}
  \If{$k==0$}
    {\bf call} localStrassen($A$, $B$, $C$);
    {\bf return}
  \EndIf
  \For{$i = 1\ldots 7$}
    \State locally compute $S_i$ and $T_i$ from $A$ and $B$ \Statex\Comment{following Strassen's algorithm}
  \EndFor
  \For{$i = 1\ldots 7$}
    \State target = rank
    \State target$[k]$ = i
    \State send $S_i$ to target
    \State receive into $L$ \Statex\Comment{One part of $L$ comes from each of 7 processors}
    \State send $T_i$ to target
    \State receive into $R$ \Statex\Comment{One part of $R$ comes from each of 7 processors}
  \EndFor
  \State {\bf call} BFS($L$, $R$, $P$, $k-1$, rank )
  \For{$i = 1\ldots 7$}
    \State target = rank
    \State target[$k$] = $i$
    \State send $i^{\mathrm{th}}$ part of $P$ to target
    \State receive from target into $Q_i$
  \EndFor
  \State locally compute $C$ from $Q_i$ \Statex\Comment{following Strassen's algorithm}
\EndProcedure
\end{algorithmic}
\end{algorithm}

\subsection{Unlimited Memory scheme}

In the UM scheme, we take $k=\log_7 P$ BFS steps in a row.  Since a BFS step reduces the number of processors involved in each subproblem by a factor of 7, after \(k\) BFS steps each subproblem is assigned to a single processor, and so is computed locally with no further communication costs.  We first describe a BFS step in more detail.

The matrices $A$ and $B$ are initially distributed as described in Section~\ref{sec:datalayout}.  In order to take a recursive step, the 14 matrices $S_1, \dots S_7, T_1, \dots, T_7$ must be computed.  Each processor allocates space for all 14 matrices and performs local additions and subtractions to compute its portion of the matrices.  Recall that the submatrices are distributed identically, so this step requires no communication.  If the layouts of \(A\) and \(B\) have parameters \((n,P,s)\), the $S_i$ and the $T_i$ now have layout parameters \((n/2,P,s-1)\).

The next step is to redistribute these 14 matrices so that the 7 pairs of matrices $(S_i,T_i)$ exist on disjoint sets of $P/7$ processors.  This requires disjoint sets of $7$ processors performing an all-to-all communication step (each processor must send and receieve a message from each of the other 6).  To see this, consider the numbering of the processors base-7.  On the \(m^\textrm{th}\) BFS step, the communication is between the seven processors whose numbers agree on all digits except the \(m^\textrm{th}\) (counting from the right).  After the \(m^\textrm{th}\) BFS step, the set of processors working on a given subproblem share the same \(m\)-digit suffix.  After the above communication is performed, the layout of \(S_i\) and \(T_i\) has parameters \((n/2,P/7,s-1)\), and the sets of processors that own the \(T_i\) and \(S_i\) are disjoint for different values of \(i\).  Note that since each all-to-all only involves seven processors no matter how large \(P\) is, this algorithm does not have the scalability issues that typically come from an all-to-all communication pattern.

\subsubsection{Memory requirements}

The extra memory required to take one BFS step is the space to store all $7$ triples $S_j$, $T_j$, $Q_j$.  Since each of those matrices is $\frac14$ the size of $A$, $B$, and $C$, the extra space required at a given step is $7/4$ the extra space required for the previous step.  We assume that no extra memory is required for the local multiplications.\footnote{If one does not overwrite the input, it is impossible to run Strassen in place; however using a few temporary matrices affects the analysis here by a constant factor only.}  Thus, the total local memory requirement for taking $k$ BFS steps is given by
\begin{align*}
\text{Mem}_\text{UM}(n,P) &= \frac{3n^2}{P} \sum_{i=0}^{k} \lt(\frac74\rt)^i=\frac{7n^2}{P^{2/{\omega_0}}} - \frac{4n^2}{P}\\
& = \Theta\lt(\frac{n^2}{P^{2/\omega_0}}\rt).
\end{align*}

\subsubsection{Computational costs}

The computation required at a given BFS step is that of the local additions and subtractions associated with computing the $S_i$ and $T_i$ and updating the output matrix $C$ with the $Q_i$.  Since Strassen performs 18 additions and subtractions, the computational cost recurrence is
$$F_\text{UM}(n,P)=18\lt(\frac{n^2}{4P} \rt)+F_\text{UM}\lt(\frac n2,\frac P7\rt)$$
with base case $F_\text{UM}(n,1)=c_sn^{\omega_0}-6n^2$, where \(c_s\) is the constant of Strassen's algorithm.  See Appendix~\ref{app:str-win} for more details.  The solution to this recurrence is
$$F_\text{UM}(n,P)=\frac{c_sn^{\omega_0}-6n^2}{P} = \Theta\lt(\frac{n^{\omega_0}}{P}\rt).$$

\subsubsection{Communication costs}

Consider the communication costs associated with the UM scheme.  Given that the redistribution within a BFS step is performed by an all-to-all communication step among sets of $7$ processors, each processor sends $6$ messages and receives $6$ messages to redistribute $S_1,\dots, S_7$, and the same for $T_1,\dots, T_7$.  After the products $Q_i=S_iT_i$ are computed, each processor sends $6$ messages and receive $6$ messages to redistribute $Q_1, \dots, Q_7$.  The size of each message varies according to the recursion depth, and is the number of words a processor owns of any $S_i$, $T_i$, or $Q_i$, namely $\frac{n^2}{4P}$ words.

As each of the $Q_i$ is computed simultaneously on disjoint sets of $P/7$ processors, we obtain a cost recurrence for the entire UM scheme:
\begin{align*}
BW_\text{UM}(n,P)&=36\frac{n^2}{4P}+BW_\text{UM}\lt(\frac n2,\frac P7\rt)\\
L_\text{UM}(n,P)&=36 +L_\text{UM}\lt(\frac n2,\frac P7\rt)
\end{align*}
with base case $L_\text{UM}(n,1)=BW_\text{UM}(n,1)=0$.  Thus
\begin{align}
  BW_\text{UM}(n,P)&=\frac{12n^2}{P^{2/{\omega_0}}} - \frac{12n^2}{P} = \Theta\lt(\frac{n^2}{P^{2/\omega_0}}\rt)\nonumber \\
  \label{eqn:UMcomm}
  L_\text{UM}(n,P)&=36\log_7 P =\Theta\lt(\log P\rt).
\end{align}

\subsection{Limited Memory scheme}

In this section we discuss a scheme for traversing Strassen's recursion tree in the context of limited memory.  In the LM scheme, we take $\ell$ DFS steps in a row followed by $k$ BFS steps in a row, where $\ell$ is minimized subject to the memory constraints.  That is, we use a sequence of DFS steps to reduce the problem size so that we can use the UM scheme on each subproblem without exceeding the available memory. 

Consider taking a single DFS step.  Rather than allocating space for and computing all 14 matrices $S_1, T_1, \dots, S_7, T_7$ at once, the DFS step requires allocation of only one subproblem, and each of the $Q_i$ will be computed in sequence.  

Consider the $i^\text{th}$ subproblem: as before, both $S_i$ and $T_i$ can be computed locally.  After $Q_i$ is computed, it is used to update the corresponding quadrants of $C$ and then discarded so that its space in memory  (as well as the space for $S_i$ and $T_i$) can be re-used for the next subproblem.  In a DFS step, no redistribution occurs.  After $S_i$ and $T_i$ are computed, all processors participate in the computation of $Q_i$.

We assume that some extra memory is available.  To be precise, assume the matrices $A$, $B$, and $C$ require only $\frac13$ of the available memory:
\begin{equation}
\label{eqn:lm}
\frac{3n^2}{P}\leq \frac13 M.
\end{equation}

In the LM scheme, we set
\begin{equation}
\label{eqn:ell}
\ell = \max\lt\{0,\left\lceil\log_2 \frac{4n}{P^{1/{\omega_0}}M^{1/2}}\right\rceil\rt\}.
\end{equation}

The following subsection shows that this choice of $\ell$ is sufficient not to exceed the memory capacity.

\subsubsection{Memory requirements}

The extra memory requirement for a DFS step is the space to store one subproblem.  Thus, the extra space required at this step is $1/4$ the space required to store $A$, $B$, and $C$.  The local memory requirements for the LM scheme is given by
\begin{align*}
\text{Mem}_\text{LM}(n,P) &= \frac{3n^2}{P}\sum_{i=0}^{\ell-1}\lt(\frac14\rt)^i + \text{Mem}_\text{UM}\lt(\frac{n}{2^\ell},P\rt) \\
&\leq \frac M3\sum_{i=0}^{\ell-1}\lt(\frac14\rt)^i+\frac{7\lt(\frac{n}{2^\ell}\rt)^2}{P^{2/{\omega_0}}} \\
&\leq \frac{127}{144} M<M,
\end{align*}
where the last line follows from \eqref{eqn:ell} and \eqref{eqn:lm}.  Thus, the limited memory scheme does not exceed the available memory.

\subsubsection{Computational costs}

As in the UM case, the computation required at a given DFS step is that of the local additions and subtractions associated with computing the $S_i$ and $T_i$ and updating the output matrix $C$ with the $Q_i$.  However, since all processors participate in each subproblem and the subproblems are computed in sequence, the recurrence is given by
$$F_\text{LM}(n,P)=18\lt(\frac{n^2}{4P} \rt)+7\cdot F_\text{LM}\lt(\frac n2,P\rt).$$
After $\ell$ steps of DFS, the size of a subproblems is $\frac{n}{2^\ell}\times\frac{n}{2^\ell}$, and there are $P$ processors involved.  We take $k$ BFS steps to compute each of these \(7^\ell\) subproblems.  Thus $$F_\text{LM}\lt(\frac{n}{2^\ell},P\rt) = F_\text{UM}\lt(\frac{n}{2^\ell},P\rt),$$
and
\begin{align*}
F_\text{LM}\lt(n,P\rt) &= \frac{18n^2}{4P} \sum_{i=0}^{\ell-1} \lt(\frac74\rt)^i + 7^\ell \cdot F_\text{UM}\lt(\frac{n}{2^\ell},P\rt) \\
&= \frac{c_sn^{\omega_0}-6n^2}{P}=\Theta\lt(\frac{n^\omega_0}{P}\rt).
\end{align*}

\subsubsection{Communication costs}

Since there are no communication costs associated with a DFS step, the recurrence is simply 
\begin{align*}
BW_\text{LM}(n,P)&=7\cdot BW_\text{LM}\lt(\frac n2,P\rt)\\
L_\text{LM}(n,P)&=7\cdot L_\text{LM}\lt(\frac n2,P\rt)
\end{align*}
 with base cases
\begin{align*}
BW_\text{LM}\lt(\frac{n}{2^\ell},P\rt) &= BW_\text{UM}\lt(\frac{n}{2^\ell},P\rt)\\
L_\text{LM}\lt(\frac{n}{2^\ell},P\rt) &= L_\text{UM}\lt(\frac{n}{2^\ell},P\rt).
\end{align*}
Thus the total communication costs are given by
\begin{align}
BW_\text{LM}\lt(n,P\rt) &= 7^\ell \cdot BW_\text{UM}\lt(\frac{n}{2^\ell},P\rt) \nonumber\\
 &\leq \frac{12\cdot 4^{{\omega_0}-2}n^{\omega_0}}{PM^{{\omega_0}/2-1}} \nonumber\\
&=\Theta\lt(\frac{n^{\omega_0}}{PM^{\omega_0/2-1}}\rt).\nonumber\\
 L_\text{LM}\lt(n,P\rt) &= 7^\ell \cdot L_\text{UM}\lt(\frac{n}{2^\ell},P\rt)\nonumber\\
&\leq \frac{(4n)^{\omega_0}}{PM^{{\omega_0}/2}} 36 \log_7 P \nonumber\\
&=\Theta\lt(\frac{n^{\omega_0}}{PM^{\omega_0/2}}\log P\rt). \label{eqn:LMcomm}
\end{align}

\subsection{Communication optimality}
\label{sec:algorithm:optimal}
\begin{proof} ({\em of Theorem~\ref{thm:main2}}).  
In the case that \(M\geq \text{Mem}_\text{UM}(n,P)=\Omega\lt(\frac{n^2}{P^{2/\omega_0}}\rt)\) the UM scheme is possible.  Then the communication costs are given by \eqref{eqn:UMcomm} which matches the lower bound of Theorem~\ref{thm:mem-indep-lb}.  Thus the UM scheme is communication-optimal (up to a logarithmic factor in the latency cost and assuming that the data is initially distributed as described in Section~\ref{sec:datalayout}).  For smaller values of \(M\), the LM scheme must be used.  Then the communication costs are given by \eqref{eqn:LMcomm} and match the lower bound of Theorem~\ref{thm:mem-dep-lb}, so the LM scheme is also communication-optimal.
\end{proof}

We note that for the LM scheme, since both the computational and communication costs are proportional to $\frac1P$, we can expect perfect strong scaling: given a fixed problem size, increasing the number of processors by some factor will decrease each cost by the same factor.  However, this strong scaling property has a limited range.  As \(P\) increases, holding everything else constant, the global memory size \(PM\) increases as well.  The limit of perfect strong scaling is exactly when there is enough memory for the UM scheme.  See \cite{BallardDemmelHoltzLipshitzSchwartz12a} for details.

\section{Analysis of Other Algorithms}
\label{sec:otheralg}
In the section we detail the asymptotic communication costs of other matrix multiplication algorithms, both classical and Strassen-based.  These communication costs and the corresponding lower bounds are summarized in Table~\ref{tbl:summary}.

Many of the algorithms described in this section are hybrids of two different algorithms.  We use the convention that the names of the hybrid algorithms are composed of the names of the two component algorithms, hyphenated.  The first name describes the algorithm used at the top level, on the largest problems, and the second describes the algorithm used at the base level on smaller problems.

\label{sec:parallelalgforStrassen}
\begin{table*}[!ht]
\begin{center}
  \setlength{\extrarowheight}{.5em}
  \begin{tabular}{| c | r | c | c | c |}
    \cline{3-5}
    \multicolumn{1}{c}{}& & \textbf{Flops} & \textbf{Bandwidth} & \textbf{Latency} \\ \cline{3-5}\multicolumn{1}{c}{}\\[-1.4em]\hline
    \multirow{4}{*}{\rotatebox{90}{Classical\hskip .6cm}} & Lower Bound \cite{IronyToledoTiskin04} &\T\B $\frac{n^3}{P}$ & $\max \lt\{\frac{n^3}{PM^{1/2}},\frac{n^2}{P^{2/3}}\rt\}$ & $\max\lt\{\frac{n^3}{PM^{3/2}},1\rt\}$ \\[.5em] \cline{2-5}
    & 2D \cite{Cannon69,SUMMA} & $\frac{n^3}{P}$ & $\frac{n^2}{P^{1/2}}$ & $P^{1/2}$ \\[.3em] \cline{2-5}
    & 3D \cite{AgarwalBalleGustavsonJoshiPalkar95,Berntsen89} & $\frac{n^3}{P}$ & $\frac{n^2}{P^{2/3}}$ & $\log P$\\[.3em]\cline{2-5}
    & 2.5D (optimal) \cite{SolomonikDemmel11}& $\frac{n^3}{P}$ & $\max \lt\{\frac{n^3}{PM^{1/2}},\frac{n^2}{P^{2/3}}\rt\}$ & $\frac{n^3}{PM^{3/2}}+\log P$ \\[.5em] \hline\hline
    \multirow{6}{*}{\rotatebox{90}{Strassen-based\hskip.9cm}} & Lower Bound \cite{BallardDemmelHoltzLipshitzSchwartz12a,BallardDemmelHoltzSchwartz11b}& $\frac{n^{\omega_0}}{P}$ & $\max\lt\{\frac{n^{\omega_0}}{PM^{{\omega_0}/2-1}},\frac{n^2}{P^{2/{\omega_0}}}\rt\}$ & $\max\lt\{\frac{n^{\omega_0}}{PM^{{\omega_0}/2}},1\rt\}$ \\[.5em] \cline{2-5}
    & 2D-Strassen \cite{LD95}& $\frac{n^{\omega_0}}{P^{({\omega_0}-1)/2}}$ & $\frac{n^2}{P^{1/2}}$ & $P^{1/2}$ \\[.3em] \cline{2-5}
    & Strassen-2D \cite{GSvdG95,LD95}& \T\B$\lt(\frac78\rt)^\ell\frac{n^3}{P}$ & $\lt(\frac74\rt)^\ell\frac{n^2}{P^{1/2}}$ & $7^\ell P^{1/2}$\\[.3em] \cline{2-5}
    & 2.5D-Strassen & $\max\lt\{\frac{n^3}{PM^{3/2-\omega_0/2}},\frac{n^{\omega_0}}{P^{\omega_0/3}}\rt\}$ & $\max\lt\{\frac{n^3}{PM^{1/2}},\frac{n^2}{P^{2/3}}\rt\}$ & $\frac{n^3}{PM^{3/2}}+\log P$\\[.5em] \cline{2-5}
    & Strassen-2.5D & $\lt(\frac78\rt)^\ell\frac{n^3}{P}$ & $\max\lt\{\lt(\frac78\rt)^\ell\frac{n^3}{PM^{1/2}},\lt(\frac74\rt)^\ell\frac{n^2}{P^{2/3}}\rt\}$ & $\lt(\frac 78\rt)^\ell\frac{n^3}{PM^{3/2}}+7^\ell\log P$\\[.5em] \cline{2-5}
    & \bf CAPS (optimal)& $\frac{n^{\omega_0}}{P}$ & $\max\lt\{\frac{n^{\omega_0}}{PM^{{\omega_0}/2-1}},\frac{n^2}{P^{2/{\omega_0}}}\rt\}$ & $\max\lt\{\frac{n^{\omega_0}}{PM^{{\omega_0}/2}}\log P,\log P\rt\}$\\[.5em] \hline
  \end{tabular}
\end{center}
\caption{Asymptotic matrix multiplication computational and communication costs of algorithms and corresponding lower bounds. Here \(\omega_0=\log_2 7\approx 2.81\) is the exponent of Strassen; \(\ell\) is the number of Strassen steps taken.  None of the Strassen-based algorithms except for CAPS attain the lower bounds of Section~\ref{sec:LB}; see Section~\ref{sec:otheralg} for a discussion of each.}
\label{tbl:summary}
\end{table*}

\subsection{Classical Algorithms}
\label{sec:classical_algs}

Classical algorithms must communicate asymptotically more than an optimal Strassen-based algorithm.  To compare the lower bounds, it is necessary to consider three cases for the memory size: when the memory-dependent bounds dominate for both classical and Strassen, when the memory-dependent bound dominates for classical, but the memory-independent bound dominates for Strassen, and when the memory-independent bounds dominate for both classical and Strassen.  This analysis is detailed in Appendix~\ref{app:ratio}.  Briefly, the factor by which the classical bandwidth cost exceeds the Strassen bandwidth cost is \(P^a\) where \(a\) ranges from \(\frac{2}{\omega_0}-\frac23\approx 0.046\) to \(\frac{3-\omega_0}{2}\approx 0.10\) depending on the relative problem size.  The same sort of analysis is used throughout Section~\ref{sec:otheralg} to compare each algorithm with the Strassen-based lower bounds.

Various parallel classical matrix multiplication algorithms minimize communication relative to the classical lower bounds for certain amounts of local memory \(M\).  For example, Cannon's algorithm \cite{Cannon69} minimizes communication for \(M=O(n^2/P)\).  Several more practical algorithms exist (such as SUMMA \cite{SUMMA}) which use the same amount of local memory and have the same asymptotic communication costs.  We call this class of algorithms ``2D'' because the communication patterns follow a two-dimensional processor grid.

Another class of algorithms, known as ``3D"  \cite{Berntsen89,AgarwalBalleGustavsonJoshiPalkar95} because the communication pattern maps to a three-dimensional processor grid, uses more local memory and reduces communication relative to 2D algorithms.  This class of algorithms minimizes communication relative to the classical lower bounds for \(M=\Omega(n^2/P^{2/3})\).  As shown in \cite{BallardDemmelHoltzLipshitzSchwartz12a}, it is not possible to use more memory than \(M=\Theta(n^2/P^{2/3})\) to reduce communication.  

Recently, a more general algorithm has been developed which minimizes communication in all cases.  Because it reduces to a 2D and 3D for the extreme values of \(M\) but interpolates for the values between, it is known as the ``2.5D'' algorithm \cite{SolomonikDemmel11}.

\subsection{2D-Strassen}
\label{sec:alg:can-str}

One idea to parallelize Strassen-based algorithms is to use a 2D classical algorithm for the inter-processor communication, and use the fast matrix multiplication algorithm locally \cite{LD95}.  We call such an algorithm ``2D-Strassen''.  It is straightforward to implement, but cannot attain all the computational speedup from Strassen since it uses a classical algorithm for part of the computation.  In particular, it does not use Strassen for the largest matrices, when Strassen  provides the greatest reduction in computation.  As a result, the computational cost exceeds $\Theta(n^{\omega_0}/P)$ by a factor of $P^{(3-\omega_0)/2}\approx P^{0.10}$.  The 2D-Strassen algorithm has the same communication cost as 2D algorithms, and hence does not match the communication costs of CAPS.  In comparing the 2D-Strassen bandwidth cost, \(\Theta(n^2/P^{1/2})\), to the CAPS bandwidth cost in Section~\ref{sec:algorithm}, note that for the problem to fit in memory we always have \(M=\Omega(n^2/P)\).  The bandwidth cost exceeds that of CAPS by a factor of \(P^a\), where \(a\) ranges from \((3-\omega_0)/2 \approx .10\) to \(2/\omega_0-1/2 \approx .21\), depending on the relative problem size.  Similarly, the latency cost, \(\Theta(P^{1/2})\), exceeds that of CAPS by a factor of \(P^a\) where \(a\) ranges from \((3-\omega_0)/2 \approx .10\) to \(1/2 = .5\).

\subsection{Strassen-2D}
\label{sec:alg:str-can}

The ``Strassen-2D'' algorithm applies \(\ell\) DFS steps of Strassen's algorithm at the top level, and performs the \(7^\ell\) smaller matrix multiplications using a 2D algorithm.  By choosing certain data layouts as in Section~\ref{sec:datalayout}, it is possible to do the additions and subtractions for Strassen's algorithm without any communication \cite{LD95}.  However, Strassen-2D is also unable to match the communication costs of CAPS.  Moreover, the speedup of Strassen-2D in computation comes at the expense of extra communication.  For large numbers of Strassen steps \(\ell\), Strassen-2D can approach the computational lower bound of Strassen, but each step increases the bandwidth cost by a factor of \(\frac 74\) and the latency cost by a factor of \(7\).  Thus the bandwidth cost of Strassen-2D is a factor of \(\lt( \frac 74\rt)^\ell\) higher than 2D-Strassen, which is already higher than that of CAPS.  The latency cost is even worse: Strassen-2D is a factor of \(7^\ell\) higher than 2D-Strassen.

One can reduce the latency cost of Strassen-2D at the expense of a larger memory footprint.  Since Strassen-2D runs a 2D algorithm \(7^\ell\) times on the same set of processors, it is possible to pack together messages from independent matrix multiplications.  In the best case, the latency cost is reduced to the cost of 2D-Strassen, which is still above that of CAPS, at the expense of using a factor of \(\lt(\frac 74\rt)^\ell\) more memory.

\subsection{2.5D-Strassen}
\label{sec:25d-strassen}
A natural idea is to replace a 2D classical algorithm in 2D-Strassen with the superior 2.5D classical algorithm to obtain an algorithm we call 2.5D-Strassen.  This algorithm uses the 2.5D algorithm for the inter-processor communication, and then uses Strassen for the local computation.  When \(M=\Theta(n^2/P)\), 2.5D-Strassen is exactly the same as 2D-Strassen, but when there is extra memory it both decreases the communication cost and decreases the computational cost since the local matrix multiplications are performed (using Strassen) on larger matrices.  To be precise, the computational cost exceeds the lower bound by a factor of \(P^a\) where \(a\) ranges from \(1-\frac{\omega_0}{3}\approx 0.064\) to \(\frac{3-\omega_0}{2}\approx 0.10\) depending on the relative problem size.  The bandwidth cost
exceeds the bandwidth cost of CAPS by a factor of \(P^a\) where \(a\) ranges from \(\frac{2}{\omega_0}-\frac23\approx 0.046\) to \(\frac{3-\omega_0}{2}\approx 0.10\).  In terms of latency, the cost of \(\frac{n^3}{PM^{3/2}}+\log P\) exceeds the latency cost of CAPS by a factor ranging from \(\log P\) to \(P^{(3-\omega_0)/2}\approx P^{0.10}\), depending on the relative problem size.

\subsection{Strassen-2.5D}
\label{sec:strassen-25d}
Similarly, by replacing a 2D algorithm with 2.5D in Strassen-2D, one obtains the new algorithm we call Strassen-2.5D.  First one takes \(\ell\) DFS steps of Strassen, which can be done without communication, and then one applies the 2.5D algorithm to each of the \(7^\ell\) subproblems.  The computational cost is exactly the same as Strassen-2D, but the communication cost will typically be lower.  Each of the \(7^\ell\) subproblems is multiplication of \(n/2^\ell\times n/2^\ell\) matrices. Each subproblem uses only \(1/4^\ell\) as much memory as the original problem.  Thus there may be a large amount of extra memory available for each subproblem, and the lower communication costs of the 2.5D algorithm help.
The choice of \(\ell\) that minimizes the bandwidth cost is
\[\ell_\mathrm{opt}=\max\lt\{0,\lt\lceil\log_2\frac{n}{M^{1/2}P^{1/3}}\rt\rceil\rt\}.\]
The same choice minimizes the latency cost.  Note that when \(M\geq \frac{n^2}{P^{2/3}}\), taking zero Strassen steps minimizes the communication within the constraints of the Strassen-2.5D algorithm.  With \(\ell=\ell_\mathrm{opt}\), the bandwidth cost is a factor of \(P^{1-\omega_0/3}\approx P^{0.064}\) above that of CAPS.  Additionally, the computational cost is not optimal, and using \(\ell=\ell_\mathrm{opt}\), the computational cost exceeds the optimal by a factor of \(P^{1-\omega_0/3}M^{3/2-\omega_0/2}\approx P^{0.064}M^{0.096}\).

It is also possible to take \(\ell>\ell_\mathrm{opt}\) steps of Strassen to decrease the comptutational cost further.  However the decreased computational cost comes at the expense of higher communication cost, as in the case of Strassen-2D.  In particular, each extra step over \(\ell_\mathrm{opt}\) increases the bandwidth cost by a factor of \(\frac74\) and the latency cost by a factor of 7.  As with Strassen-2D, it is possible to use extra memory to pack together messages from several subproblems and decrease the latency cost, but not the bandwidth cost.

\section{Performance Results}
\label{sec:performance}
We have implemented CAPS using MPI on a Cray XT4, and compared it to various previous classical and Strassen-based algorithms.  The benchmarking data is shown in Figure~\ref{fig:performance}.
\subsection{Experimental setup}
The nodes of the Cray XT4 have 8GB of memory and a quad-core AMD ``Bupdapest'' processor running at 2.3GHz.  We treat the entire node as a single processor, and when we use the classical algorithm we call the optimized threaded BLAS in Cray's LibSci to provide parallelism between the four cores in a node.  The peak flop rate is 9.2 GFLOPS per core, or 36.8 GFLOPS per node.  The machine consists of 9,572 nodes.  All the data in Figure~\ref{fig:performance} is for multiplying two square matrices with \(n=94080\). 
\subsection{Performance}
\label{sec:performance:performance}
\begin{figure*}[!ht]
  \centering
  \includegraphics[width=7in]{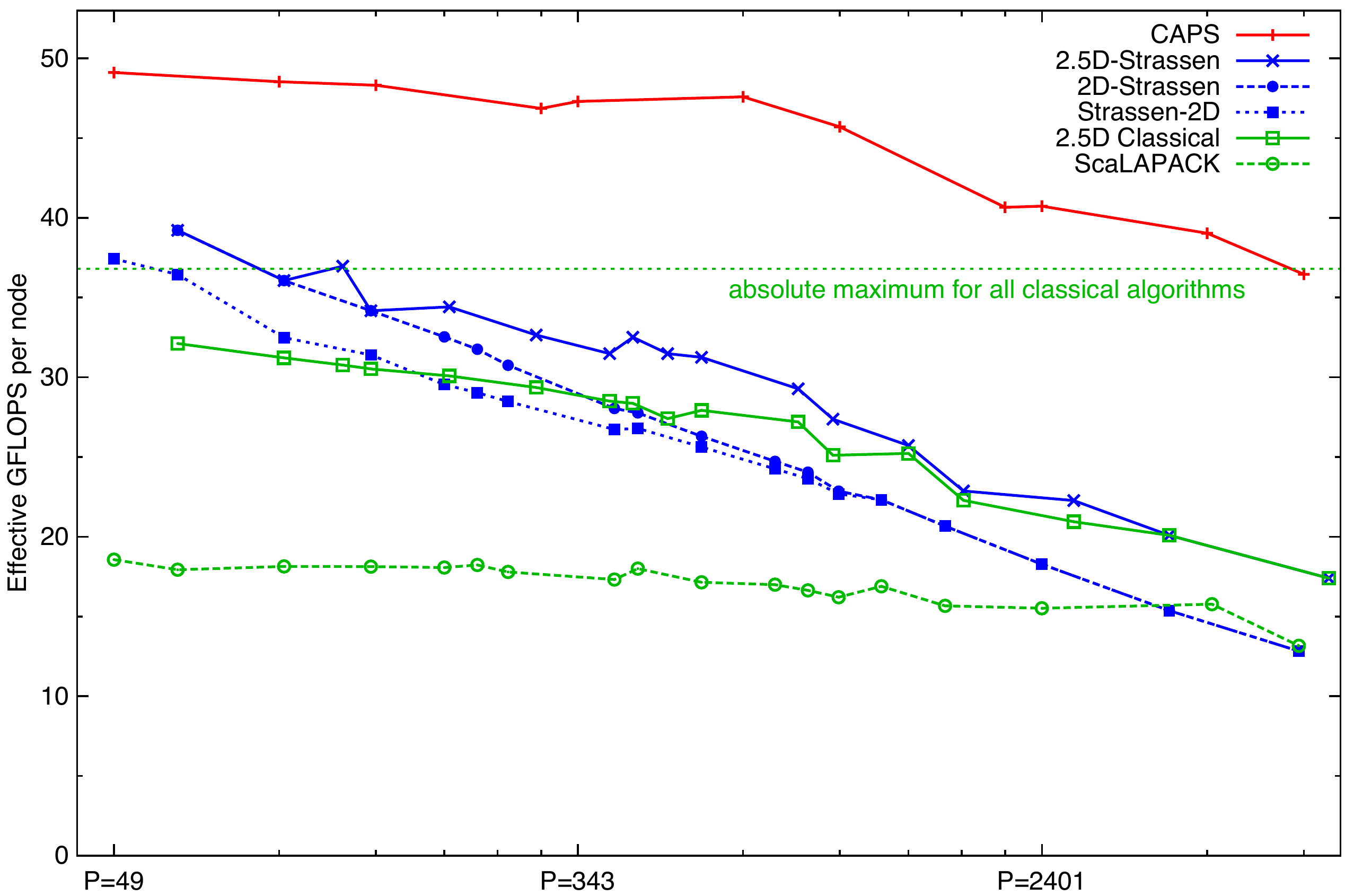}
  \caption{Strong scaling performance of various matrix multiplication algorithms on Cray XT4 for fixed problem size $n=94080$.  The top line is CAPS as described in Section~\ref{sec:algorithm}, and substantially outperforms all the other classical and Strassen-based algorithms.  The horizontal axis is the number of nodes in log-scale.  The vertical axis is effective GFLOPS, which are a performance measure rather than a flop rate, as discussed in Section~\ref{sec:performance:performance}.  See Section~\ref{sec:performance:details} for a description of each implementation.}
  \label{fig:performance}
\end{figure*}

Note that the vertical scale of Figure~\ref{fig:performance} is ``effective GFLOPS'', which is a useful measure for comparing classical and fast matrix multiplication algorithms.  It is calculated as
\begin{equation}
\textrm{Effective GFLOPS}=\frac{2n^3}{(\textrm{Execution time in seconds})10^9}.
\end{equation}
For classical algorithms, which perform \(2n^3\) floating point operations, this gives the actual GFLOPS.  For fast matrix multiplication algorithms it gives the performance relative to classical algorithms, but does not accurately represent the number of floating point operations performed.

Our algorithm outperforms all previous algorithms, and attains performance as high as 49.1 effective GFLOPS/node, which is \(33\%\) above the theoretical maximum for all classical algorithms.  Compared with the best classical implementation, our speedup ranges from \(51\%\) for small values of \(P\) up to \(94\%\) when using most of the machine.  Compared with the best previous parallel Strassen algorithms, our speedup ranges from \(24\%\) up to \(184\%\).  Unlike previous Strassen algorithms, we are able to attain substantial speedups over the entire range of processors.

\subsection{Strong scaling}
Figure~\ref{fig:performance} is a strong scaling plot: the problem size is fixed and each algorithm is run with \(P\) ranging from the minimum that provides enough memory up to the largest allowed value of \(P\) smaller than the size of the machine.  Perfect strong scaling corresponds to a horizontal line in the plot.  As the communication analysis predicts, CAPS exhibits better strong scaling than any of the other algorithms (with the exception of ScaLAPACK, which obtains very good strong scaling by having poor performance for small values of \(P\)).

\subsection{Details of the implementations}
\label{sec:performance:details}

\subsubsection{CAPS}
This implementation is the CAPS algorithm, with a few modifications from the presentation in Section~\ref{sec:algorithm}.  First, when computing locally it switches to classical matrix multiplication below some size \(n_0\).  Second, it is generalized to run on \(P=c 7^k\) processors for \(c\in\{1,2,3,6\}\) rather than just \(7^k\) processors.  As a result, the base-case classical matrix multiplication is done on \(c\) processors rather than 1.  Finally, implementation uses the Winograd variant of Strassen; see Appendix~\ref{app:str-win} for more details.  Every point in the plot is tuned to use the best interleaving pattern of BFS and DFS steps, and the best total number of Strassen steps.  For points in the figure, the optimal total number of Strassen steps is always 5 or 6.

\subsubsection{ScaLAPACK}
We use ScaLAPACK \cite{SCALAPACK} as optimized by Cray in LibSci.  This is an implementation of the SUMMA algorithm, and can run on an arbitrary number of processors.  It should give the best performance if \(P\) is a perfect square so the processors can be placed on a square 2D grid.  All the runs shown in Figure~\ref{fig:performance} are with \(P\) a perfect square.

\subsubsection{2.5D classical}
This is the code of \cite{SolomonikDemmel11}.  It places the \(P\) processors in a grid of size \(\sqrt{P/c}\times\sqrt{P/c}\times c\), and requires that \(\sqrt{P/c}\) and \(c\) are integers with \(1\leq c\leq P^{1/3}\), and \(c\) divides \(\sqrt{P/c}\).  Additionally, it gets the best performance if \(c\) is as large as possible, given the constraint that \(c\) copies of the input and output matrices fit in memory.  In the case that \(c=1\) this code is an optimized implementation of SUMMA.  The values of \(P\) and \(c\) for the runs in Figure~\ref{fig:performance} are chosen to get the best performance.

\subsubsection{Strassen-2D}
Following the algorithm of \cite{GSvdG95,LD95}, this implementation uses the DFS code from the implementation of CAPS at the top level, and then uses the optimized SUMMA code from the 2.5D implementation with \(c=1\).
Since the classical code requires that \(P\) is a perfect square, this requirement applies here as well.  The number of Strassen steps taken is tuned to give the best performance for each \(P\) value, and the optimal number varies from 0 to 2.

\subsubsection{2D-Strassen}
Following the algorithm of \cite{LD95}, the 2D-Strassen implementation is analagous to the Strassen-2D implementation, but with the classical algorithm run before taking local Strassen steps.  Similarly the same code is used for local Strassen steps here and in our implementation of CAPS.  This code also requires that \(P\) is a perfect square.  The number of Strassen steps is tuned for each \(P\) value, and the optimal number varies from  0 to 3.

\subsubsection{2.5D-Strassen}
This implementation uses the 2.5D implementation to reduce the problem to one processor, then takes several Strassen steps.  The processor requirements are the same as for the 2.5D implementation.  The number of Strassen steps is tuned for each number of processors, and the optimal number varies from 0 to 3.  We also tested the Strassen-2.5D algorithm, but its performance was always lower than 2.5D-Strassen in our experiments.

\section{Conclusions/Future Work}
\label{sec:conclusion}

\subsection{Stability of fast matrix multiplication}
CAPS has the same stability properties as sequential versions of Strassen.  For a complete discussion of the stability of fast matrix multiplication algorithms, see \cite{Higham96,DemmelDumitriuHoltzKleinberg07}.  We highlight a few main points here.  The tightest error bounds for classical matrix multiplication are component-wise: $|C-\hat C| \leq n \epsilon |A|\cdot |B|,$ where $\hat C$ is the computed result and $\epsilon$ is the machine precision.  Strassen and other fast algorithms do not satisfy component-wise bounds but do satisfy the slightly weaker norm-wise bounds: $\|C-\hat C\| \leq f(n) \epsilon \|A\| \|B\|,$ where $\|A\| = \max_{i,j} A_{ij}$ and $f$ is polynomial in $n$ \cite{Higham96}.  Accuracy can be improved with the use of diagonal scaling matrices: $D_1CD_3 = D_1AD_2\cdot D_2^{-1}BD_3$.  It is possible to choose \(D_1,D_2,D_3\) so that the error bounds satisfy either $|C_{ij}-\hat C_{ij}|\leq f(n)\epsilon\|A(i,:)\| \|B(:,j)\|$ or \(\|C-\hat C\|\leq f(n)\epsilon\||A|\cdot|B|\|\).  By scaling, the error bounds on Strassen become comparable to those of many other dense linear algebra algorithms, such as LU and QR decomposition \cite{DemmelDumitriuHoltz07}.  Thus using Strassen for the matrix multiplications in a larger computation will often not harm the stability at all.

\subsection{Hardware scaling}
Although Strassen performs asymptotically less computation and communication than classical matrix multiplication, it is more demanding on the hardware.  That is, if one wants to run matrix multiplication near the peak CPU speed, Strassen is more demanding of the memory size and communication bandwidth.  This is because the ratio of computational cost to bandwidth cost is lower for Strassen than for classical.  
From the lower bounds in Section~\ref{sec:LB}, the asymptotic ratio of computational cost to bandwidth cost is \(M^{\omega_0/2-1}\) for Strassen-based algorithms, versus \(M^{1/2}\) for classical algorithms.  This means that it is harder to run Strassen near peak than it is to run classical matrix multiplication near peak.  
In terms of the machine parameters \(\beta\) and \(\gamma\) introduced in Section~\ref{sec:model}, the condition to be able to be computation-bound is \(\gamma M^{1/2}\geq c\beta\) for classical matrix multiplication and \(\gamma M^{\omega_0/2-1}\geq c'\beta\) for Strassen.  Here \(c\) and \(c'\) are constants that depend on the constants in the communication and computational costs of classical and Strassen-based matrix multiplication.

The above inequalities may guide hardware design as long as classical and Strassen matrix multiplication are considered important computations.  They apply both to the distributed case, where \(M\) is the local memory size and \(\beta\) is the inverse network bandwidth, and to the sequential/shared-memory case where \(M\) is the cache size and \(\beta\) is the inverse memory-cache bandwidth.

\subsection{Optimizing on-node performance}

Note that although our implementation performs above the classical peak performance, it performs well below the corresponding Strassen-Winograd peak, defined by the time it takes to perform \(c_s n^{\omega_0}/P\) flops at the peak speed of each processor.  To some extent, this is because Strassen is more demanding on the hardware, as noted above.  However we have not yet analyzed whether the amount our performance is below Strassen peak can be entirely accounted for based on machine parameters.  It is also possible that a high performance shared-memory Strassen implementation might provide substantial speedups for our implementation, as well as for 2D-Strassen and 2.5D-Strassen.

\subsection{Testing on various architectures}
We have implemented and benchmarked CAPS on only one architecture, a Cray XT4.  It remains to check that it outperforms other matrix multiplication algorithms on a variety of architectures.  On some architectures it may be more important to consider the topology of the network and redesign the algorithm to minimize contention, which we have not done.

\subsection{Improvements to the algorithm}


To be practically useful, it is important to generalize the number of processors on which CAPS can run.
To attain the communication lower bounds, CAPS as presented in Section~\ref{sec:algorithm} 
must run on \(P\) a power of seven processors.
Of course, CAPS can then be run on any number of processors by simply ignoring no more than \(\frac 67\)  of them and incurring a constant factor overhead.  Thus we can run on arbitrary \(P\)
and attain the communication and computation lower bounds up to a
constant factor.  However the computation time is still dominant in
most cases, and it is preferable to attain the computation lower bound
exactly.  It is an open question whether any algorithm
can run on arbitrary \(P\), attain the computation lower bound
exactly, and attain the communication lower bound up to a constant
factor.

Moreover, although the communication costs of this algorithm match the
lower bound up to a constant factor in bandwidth, and up to a \(\log
P\) factor in latency, it is an open question to determine the optimal constant in the lower bound and perhaps provide a new algorithm that matches it exactly.  Note that in the analysis of CAPS in Section~\ref{sec:algorithm}, the constants could be slightly improved.

\subsection{Parallelizing other algorithms}
\subsubsection{Another optimal classical algorithm}
We can apply our parallelization approach to recursive classical matrix multiplication to obtain a communication-optimal algorithm.  This algorithm has the same asymptotic communication costs as the 2.5D algorithm \cite{SolomonikDemmel11}.  We observed comparable performance to the 2.5D algorithm on our experimental platform.  As with CAPS, this algorithm has not been optimized for contention, whereas the 2.5D algorithm is very well optimized for contention on torus networks.

\subsubsection{Other fast matrix multiplication algorithms}
Our approach of executing a recursive algorithm in parallel by traversing the recursion tree in DFS (sequential) or BFS (parallel) manners is not limited to Strassen's algorithm.  All fast square matrix multiplication algorithms are built out of ways to multiply \(n_0\times n_0\) matrices using \(q<n_0^3\) multiplications.  Like with Strassen and Strassen-Winograd, they compute \(q\) linear combinations of entries of each of \(A\) and \(B\), multiply these pairwise, then compute the entries of \(C\) as linear combinations of these.\footnote{By \cite{Raz03}, \emph{all} fast matrix multiplication algorithms can be expressed in this bilinear form.}  CAPS can be easily generalized to any such multiplication, with the following modifications:
\begin{itemize}
\item The number of processors \(P\) is a power of \(q\).
\item The data layout must be such that all \(n_0^2\) blocks of \(A\), \(B\), and \(C\) are distributed equally among the \(P\) processors with the same layout.
\item The BFS and DFS determine whether the \(q\)  multiplications are performed in parallel or sequentially.
\end{itemize}
The communication costs are then exactly as above, but with \(\omega_0=\log_{n_0} q\).

It is unclear whether any of the faster matrix multiplication algorithms are useful in practice.  One reason is that the fastest algorithms are not explicit.  Rather, there are non-constructive proofs that the algorithms exist.  To implement these algorithms, they would have to be found, which appears to be a difficult problem.
The generalization of CAPS described in this section does apply to all of them, so we have proved the existence of a communication-avoiding non-explicit parallel algorithm corresponding to every fast matrix multiplication algorithm.  We conjecture that the algorithms are all communication-optimal, but that is not yet proved since the lower bound proofs in \cite{BallardDemmelHoltzSchwartz11b,BallardDemmelHoltzLipshitzSchwartz12a} may not apply to all fast matrix multiplication algorithms.  In cases where the lower bounds do apply, they match the performance of the generalization of CAPS, and so they are communication-optimal.

\newpage
\bibliographystyle{abbrv}
\bibliography{bib_strassen} 
%
%

\appendix
\section{Strassen-Winograd Algorithm}
\label{app:str-win}
The Strassen-Winograd algorithm is usually preferred to Strassen's algorithm in practice since it requires fewer additions.  We use it for our implementation of CAPS.  Divide the input matrices \(A,B\) and output matrix \(C\) into 4 submatrices:
\small
\[
A=\left[\begin{array}{ll}A_{11} & A_{12}\\A_{21} & A_{22}\end{array}\right] \; 
B=\left[\begin{array}{ll}B_{11} & B_{12}\\B_{21} & B_{22}\end{array}\right]  \;
C=\left[\begin{array}{ll}C_{11} & C_{12}\\C_{21} & C_{22}\end{array}\right]
\]
\normalsize
Then form 7 linear combinations of the submatrices of each of \(A\) and \(B\), call these \(T_i\) and \(S_i\), respectively; multiply them pairwise; then form the submatrices of \(C\)  as linear combinations of these products:
\scriptsize
\[
\begin{array}{llll}
T_0=A_{11}  &
S_0=B_{11}  &
Q_0=T_0 \cdot S_0  &
U_{1\phantom{1}}=Q_0+Q_3\\
T_1=A_{12}  &
S_1=B_{21}  &
Q_1=T_1 \cdot S_1  &
U_{2\phantom{1}}=U_1+Q_4\\
T_2=A_{21}+A_{22}  &
S_2=B_{12}+B_{11}  &
Q_2=T_2 \cdot S_2  &
U_{3\phantom{1}}=U_1+Q_2\\
T_3=T_2-A_{12}  &
S_3=B_{22}-S_2  &
Q_3=T_3 \cdot S_3  &
C_{11}=Q_0+Q_1\\
T_4=A_{11}-A_{21}  &
S_4=B_{22}-B_{12}  &
Q_4=T_4 \cdot S_4  &
C_{12}=U_3+Q_5\\
T_5=A_{12}+T_3  &
S_5=B_{22}  &
Q_5=T_5 \cdot S_5  &
C_{21}=U_2-Q_6\\
T_6=A_{22}  &
S_6=S_3-B_{21}  &
Q_6=T_6 \cdot S_6  &
C_{22}=U_2+Q_2
\end{array}
\]
\normalsize

This is one step of Strassen-Winograd.  The algorithm is recursive since it can be used for each of the 7 smaller matrix multiplications.  In practice, one often uses only a few steps of Strassen-Winograd, although to attain \(O(n^{\omega_0})\) computational cost, it is necessary to recursively apply it all the way down to matrices of size \(O(1)\times O(1)\).  The precise computational cost of Strassen-Winograd is
\[\mathrm{F}(n)=c_sn^{\omega_0}-5n^2.\]
Here \(c_s\) is a constant depending on the cutoff point at which one switches to the classical algorithm.  For a cutoff size of \(n_0\), the constant is \(c_s=(2n_0+4)/n_0^{\omega_0-2}\) which is minimized at \(n_0=8\) yielding a computational cost of approximately \(3.73n^{\omega_0}-5n^2\).  If using Strassen-Winograd with cutoff \(n_0\) this should be substituted into the computational cost expressions of Section~\ref{sec:algorithm}.

\section{Communication-cost ratios}
\label{app:ratio}
In this section we derive the ratio \(R\) of classical to Strassen-based bandwidth cost lower bounds that appear in the beginning of Section~\ref{sec:otheralg}.  Note that both classical and Strassen-based lower bounds are attained by optimal algorithms.  Similar derivations apply to the other ratios quoted in that section.  Because the bandwidth cost lower bounds are different in the memory-dependent and the memory-independent cases, and the threshold between these is different for the classical and Strassen-based bounds, it is necessary to consider three cases.
\paragraph*{Case 1} \(M=\Omega(n^2/P)\) and \(M=O(n^2/P^{2/\omega_0})\).  The first condition is necessary for there to be enough memory to hold the input and output matrices; the second condition puts both classical and Strassen-based algorithms in the memory-dependent case.  Then the ratio of the bandwidth costs is:
\[R=\Theta\lt(\frac{n^3}{PM^{1/2}}\lt/\frac{n^{\omega_0}}{PM^{\omega_0/2-1}}\rt.\rt)=\Theta\lt(\lt(\frac{n^2}{M}\rt)^{(3-\omega_0)/2}\rt).\]
Using the two bounds that define this case, we obtain \(R=O(P^{(3-\omega_0)/2})\) and \(R=\Omega(P^{3/\omega_0-1})\).
\paragraph*{Case 2} \(M=\Omega(n^2/P^{2/\omega_0})\) and \(M=O(n^2/P^{2/3})\).  This means that in the classical case the memory-dependent lower bound dominates, but in the Strassen-based case the memory-independent lower bound dominates.  Then the ratio is:
\[R=\Theta\lt(\frac{n^3}{PM^{1/2}}\lt/\frac{n^2}{P^{2/\omega_0}}\rt.\rt)=\Theta\lt(\lt(\frac{n^2}{M}\rt)^{1/2}P^{2/\omega_0-1}\rt).\]
Using the two bounds that define this case, we obtain \(R=O(P^{3/\omega_0-1})\) and \(R=\Omega(P^{2/\omega_0-2/3})\).
\paragraph*{Case 3} \(M=O(P^{2/3})\).  This means that both the classical and Strassen-based lower bounds are dominated by the memory-independent cases.  Then the ratio is:
\[R=\Theta\lt(\frac{n^2}{P^{2/3}}\lt/\frac{n^2}{P^{2/\omega_0}}\rt.\rt)=\Theta\lt(P^{2/\omega_0-2/3}\rt).\]


Overall, depending on the ratio of the problem size to the available memory, the factor by which the classical bandwidth costs exceed the Strassen-based bandwidth costs is between \(\Theta(P^{2/\omega_0-2/3})\) and \(\Theta(P^{(3-\omega_0)/2})\).

\balancecolumns 
\end{document}